\documentclass[Royal,sageh,times]{sagej_hark}

\usepackage{moreverb,url}

\usepackage{hyperref}

\def\volumeyear{2020}

\def\utw{\smash{\rlap{\lower5pt\hbox{$\sim$}}}}
\def\udtw{\smash{\rlap{\lower6pt\hbox{$\approx$}}}}

\def\fdg{\hbox{$.\!\!^\circ$}}

\def\farcs{\hbox{$.\!\!^{\prime\prime}$}}

\newcommand{\rev}{}

\begin{document}

\runninghead{Petrovay}

\title{The Determination of Stellar Temperatures from Baron B. Hark\'anyi
to the Gaia Mission}

\author{Krist\'of Petrovay}

\affiliation{E\"otv\"os Lor\'and University, Budapest, Hungary}

\corrauth{Krist\'of Petrovay, E\"otv\"os Lor\'and University,
Department of Astronomy, Budapest, P\'azm\'any P.~s.~1/A, H-1117 Hungary.}
\email{K.Petrovay@astro.elte.hu}

\begin{abstract}
The first determination of the surface temperature of stars other than the Sun
is due to the Hungarian astrophysicist B\'ela Hark\'anyi. Prompted by the recent
unprecedented increase in the availability of stellar temperature estimates from
Gaia, coinciding with the 150th anniversary of Hark\'anyi's birth, this article
presents the life and work of this neglected, yet remarkable figure in the
context of the history of stellar astrophysics. 
\end{abstract}

%, K\"ovesligethy, Vogel}

\maketitle

\section{Introduction}

The second data release of the Gaia mission published in 2018 has provided
effective temperature estimates for more than 160 million
stars.\endnote{R. Andrae et~al., ``Gaia Data Release 2. First stellar 
parameters from Apsis'', {\it Astronomy \& Astrophysics} 616, 2019: id.~A8} For 
nearly 77 million stars these data are also accompanied by luminosity and radius 
values.
The release of this vast dataset provides an opportunity to recall a forgotten
episode in the history of astrophysics: the  first determination of the surface
temperatures of specific stars other than the Sun by the Hungarian
astrophysicist B\'ela Hark\'anyi in 1902.\endnote{B. Hark\'anyi, ``\"Uber die 
Temperaturbestimmung der Fixsterne auf spectralphotometrischem Wege'', 
{\it Astronomische Nachrichten} 158, 1902: pp.~17--24} As we will see, 
Hark\'anyi's approach
allows interesting parallels with the new Gaia results, including the
limitations affecting both methods. The parallels do not end here, as in a
followup study published in 1910 Hark\'anyi also pioneered the determination 
of stellar radii for single stars based on the radiation energetics 
method.\endnote{B. Hark\'anyi, ``\"Uber die Fl\"achenhelligkeit, photometrische 
Gr\"o{\ss}e und Temperatur der Sterne'', {\it Astronomische Nachrichten} 185,
1910: pp.~33--48;\\ 
Hark\'anyi B., ``Darstellung der photometrischen und  photographischen
Gr\"o{\ss}e als Funktion der Temperatur der Sterne'',  {\it Astronomische
Nachrichten} 186, 1910: pp.~161--176}

Before focusing on Hark\'anyi's work it may be in order to draw up the
wider context of how astrophysics went ``from rags to riches'' in the
past two centuries. As we are often reminded, in the early 19th
century {\rev the study of the physical constitution, chemical
composition or origin of stars and other celestial bodies ---the field
that later came to be known as astrophysics---} was widely derided as
a purely speculative enterprise without firm empirical or theoretical
basis, the only subject worthy of study for ``real'' or ``precision''
astronomy being the positions and orbits of celestial bodies, based on
highly accurate measurements and computations. This view is often
attributed to Auguste Comte, a forerunner of positivist philosophy;
however, as recently pointed out by A. H. Batten, the same opinion was
widely held among astronomers of the day like Bessel, Gauss or
Airy.\endnote{See discussion in D. B. Herrmann, {\it Geschichte der
Astronomie von Herschel bis Hertzsprung} (Berlin: VEB Deutsher Verlag
der Wissenschaften, 1975), pp. 97--99, and, more recently by A. H.
Batten, ``Comte, Mach, Planck, and Eddington: a study of influence
across generations'', {\it Journal of Astronomical History and
Heritage,} 19(1), 2016: pp.~51--60.}

The kind of hard data, precise numerical measurements related to the
constitution and physical state of stars upon which astrophysics was
ultimately to be founded as an exact science, however, became
available in 1817, the year that may, {\rev in retrospect,} well be
considered the birth year of astrophysics. {\rev (Note, however, that
the term {\it astrophysics} came into use in the second half of the
19th century only,)}. It was in this year that the German optician
Fraunhofer published his observation of a large number of sharp dark
lines in the solar spectrum, each with its well defined wavelength and
strength. While neither Fraunhofer nor any of his contemporaries knew
how to interpret these findings, it {\rev must have been} clear that
the origin of the lines has to do with, and therefore must encode
information about the physical and chemical conditions at the source,
i.e. the solar surface.

The emancipation of astrophysics took a long time, proceeding in parallel with 
the painfully slow process of decoding the information in spectra. The
first milestone along this road was the establishment of Kirchhoff's laws of
spectroscopy in 1860. While laboratory studies of the strengths of a spectral
line (Mg $\lambda$4480) in gases at varying temperatures, in comparison with
stellar spectra  allowed J. Scheiner\endnote{J. Scheiner, ``Temperature on the 
surface of fixed stars and of the Sun, compared with that of terrestrial heat 
sources'', {\it Publications of the  Astronomical Society of the Pacific} 6,
1894: pp.~172--173} to roughly bracket stellar
temperatures in the range 3,000--15,000 K, a more exact determination of
temperatures from the properties of lines in individual stellar spectra only
became possible after 1910.  Up to the first decade of the 20th century studies
of the continuous spectra offered a more accessible route to the determination
of stellar temperatures. (The process has been reviewed in more detail by
D.~H.~DeVorkin and R.~Kenat\endnote{D.~H.~DeVorkin and R. Kenat, ``Quantum 
physics and the stars (I): The establishment of a stellar temperature scale'', 
{\it Journal for the History of Astronomy} 14, 1983: pp.~102--132}.)

Pioneering work on the {\rev physics} of continuous spectra was already done in the
1870s by J. Stefan at the University of Vienna. After empirically setting up his
famous $T^4$ law, Stefan immediately went on to determine the solar effective
temperature as $6000\,$K. \endnote{J. Stefan, ``\"Uber die Beziehung
zwischen der W\"armstrahlung und der Temperatur'', {\it Sitzungsberichte
d.~Akad.~d.~Wiss.~in Wien} 79, 2, 391--428 (1879)} Yet for another 15 years
Stefan's result was generally given no more credit than other competing
radiation laws, resulting in widely differing values of the solar temperature
ranging from $1500\,$K to $10^7$ K (!). Stefan's results were finally
independently empirically confirmed in 1894 by Wilson and
Gray\endnote{W.~E.~Wilson and P.~E.~Gray, ``Experimental Investigations on the
Effective Temperature of the Sun'', {\it Proceedings of the Royal Society of 
London Series I} 55, 1894: pp.~250--251}, and were placed on a firm theoretical
basis by the spectral theories of Wien (1897) and Planck (1901), so by the end
of the 19th century the solar temperature was finally considered well
established.\endnote{J. Scheiner, ``The Temperature of the Sun. I'',  
{\it Publications of the Astronomical Society of the Pacific,} 10, 1898: 
pp. 167--179}

The first attempt at a theoretical derivation of the blackbody radiation law is
due to Stefan's student R. K\"ovesligethy.\endnote{H. Kangro, 
{\it Vorgeschichte des Planckschen Strahlungsgesetzes} 
(Wiesbaden, Franz Steiner, 1970);\\ 
L. G. Bal\'azs, M. Vargha, and E. Zsoldos,
``Rad\'o K\"ovesligethy's spectroscopic work'',
{\it Journal of Astronomical History and Heritage,} 11, 2008: pp. 124--133
}
While K\"ovesligethy's spectral theory\endnote{The theory was published in
Hungarian in 1885, while five years later it became part of K\"ovesligethy's
comprehensive German monograph on spectral theory:\\
R. K\"ovesligethy, ``A folytonos spektrumok elm\'elete''. 
{\it \'Ertekez\'esek a mathematikai tudom\'anyok k\"or\'eb\H ol} 12, 1885: 11 --  
\url{http://real-eod.mtak.hu/1622/1/Matematikai_ertekezesek_12_11.pdf};\\
R. von K\"ovesligethy, {\it Grundz\"uge einer theoretischen Spectralanalyse,}
(Halle,  H. W. Schmidt, 1890) -- 
\url{https://archive.org/details/grundzgeeinerth00kvgoog/page/n8}}
was logically coherent, it was based on false premises (amounting to a
non-relativistic aether theory), and after his PhD work, started in 1880, was
published in 1885 his theory quickly became obsolete in the light of the newly
developed Maxwellian electrodynamics. Nevertheless, K\"ovesligethy's suggested
blackbody radiation formula correctly predicted (and on sound physical grounds)
the displacement law $\lambda_{\rm max}T=\,$const., later derived by
Wien on general thermodynamic grounds. The importance of this detail from the
point of view of our subject lies in the fact that following his PhD
K\"ovesligethy obtained a teaching position at the University of Budapest where,
as a young Privatdozent, he soon encountered his first disciple: a young man
called B\'ela Hark\'anyi.

\begin{figure}
\begin{center}
\includegraphics[height=0.25\textheight]{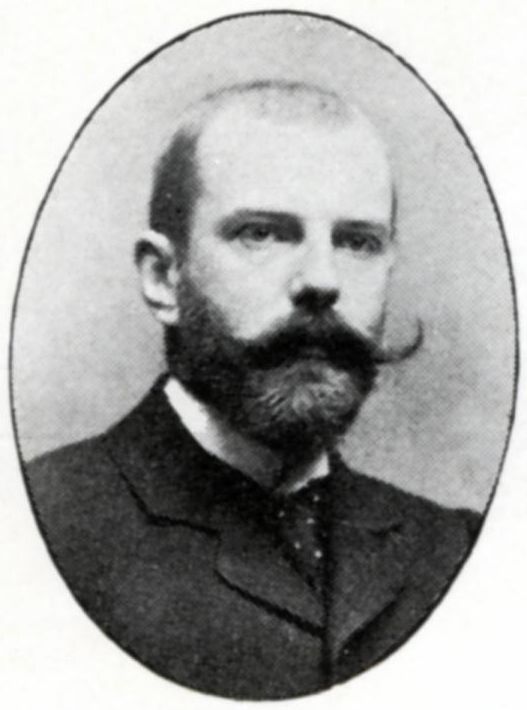}\hskip 1 cm
\includegraphics[height=0.25\textheight]{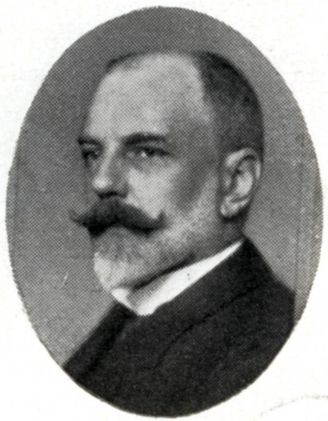}
\end{center}
\caption{Portraits of Hark\'anyi: as an observer at \'Ogyalla (left) and as a
senior Privatdozent (right).}
\label{fig:portraits}
\end{figure}

\section{B\'ela Hark\'anyi: biographical data}

B\'ela Hark\'anyi was born in 1869 in Budapest, in a wealthy landowner family of
Jewish origin. Promoted to the ranks of nobility just two years before
Hark\'anyi's birth, in 1895 they were also awarded the
title of Baron, which Hark\'anyi proudly wore and indicated on most of his
publications.\endnote{{\rev Hark\'anyi's father K\'aroly and his cousin
J\'anos were both active politicians; J\'anos Hark\'anyi served as
Minister of Commerce in the period 1913--17. In 1947, on the eve of
the Communist collectivization, the family still possessed over 22
thousand acres of land.\\
J. Gudenus, L. Szentirmay: \"Osszet\"ort c\'{\i}merek 
(Budapest, Piremon-Mozaik, 1989)\\}
J.~J.~Gudenus, ``A magyarorsz\'agi f{\H o}nemess\'eg 
XX.~sz\'azadi geneal\'ogi\'aja'', vol.~1 (Budapest, Magyar Mez\H ogazdas\'agi 
Kiad\'o, 1990);\\
B. Kempelen, ``Magyar nemes csal\'adok'', vol.~4 (Budapest, 
Grill K\'aroly K\"onyvkiad\'ov\'allalata, 1911) -- \\
\url{https://www.arcanum.hu/hu/online-kiadvanyok/Kempelen-kempelen-bela-magyar-nemes-csaladok-1/4-kotet-56A5/harkanyi-taktaharkanyi-baro-7ACB
}} 
With this family background, he never knew financial need, nor did
he ever depend on a salary. The study of astronomy was for him a gentleman's
passion, yet he pursued it with a highly devoted and professional attitude.

After studying mathematics and physics for three years at the University of
Sciences in Budapest, in 1891 he moved to Leipzig for the fourth, final year of
his BSc studies, and then to Strassburg (at the time part of Germany) for three
years of PhD studies. Finally, in 1896 he presented his PhD thesis in Budapest,
written under the guidance of K\"ovesligethy. Two ``postdoc'' years followed in
the Observatoire de Paris, then half a year in Potsdam ---it was Hark\'anyi's
financial self-reliance that allowed him to follow such a strikingly modern
career path in an age when postdoctoral positions did not yet exist. Finally, in
1899 he returned to Hungary where he was offered an observer's position in the
Astrophysical Observatory of \'Ogyalla. 

The observatory had recently been donated to the Hungarian state by
its founder, Mikl\'os Konkoly Thege, and Hark\'anyi's former professor
K\"ovesligethy was appointed to serve as vice-director under Konkoly.
In the four years spent at \'Ogyalla the two colleagues had no doubt
ample opportunity to discuss K\"ovesligethy's long-time interest, the
theory of continuous spectra, where exciting new developments took
place in these years. This led to the publication of Hark\'anyi's
seminal paper on the determination of stellar temperatures, to be
discussed below.

In 1903 both Hark\'anyi and K\"ovesligethy left \'Ogyalla and moved definitively
to Budapest where K\"ovesligethy had maintained his professor's position all along
and, from 1907, Hark\'anyi also obtained a position as
Privatdozent.\endnote{{\rev In the following year, 1908, Hark\'anyi
married Blanka Hieronymi, and in 1911 their son K\'aroly was born.
Tragically, soon after this Blanka died, and the son followed her
mother in 1921 (J. J. Gudenus, op.cit.)}} 
He spent the
rest of his life on the university cathedra. In 1911 he became a corresponding
member of the Hungarian Academy of Sciences. In normal times this should have
been followed by full professorship; these, though, were not normal times. To
Hark\'anyi's bad luck, his promotion to full professorship took place under the
short-lived Communist dictatorship of 1919, in the wake of the lost war. In the
ensuing right-wing regime all provisions of the Communist dictatorship were
declared void.\endnote{M. Vargha, ``K\"ovesligethy Rad\'o \'eletrajz\'aval
kapcsolatos dokumentumok 22.: Igazoltat\'asok a Magyar Tudom\'anyos 
Akad\'emi\'an'', {\it Konkoly Monographs} 8, 2011: pp.~69--72 -- 
\url{https://konkoly.hu/Mitteilungen/m8.pdf}} 
Hark\'anyi was demoted and remained Privatdozent until his death 
in 1932.\endnote{A. Tass, ``Erinnerungen an B. v. Hark\'anyi'', {\it
Vierteljahrsschrift der Astronomischen Gesellschaft} 68 (4), 1932: 300;\\
R. K\"ovesligethy, ``Todesanzeige'', {\it Astron. Nachrichten} 245, 1932: 5859.}

Beyond the mere biographical facts, there are only a few recollections of
Hark\'anyi's character and attitude. These recollections corroborate the
impression of a rigorous and reclusive scientist, suggested by his
portraits\endnote{{\it Potr\"atgallerie der Astronomischen
Gesellschaft} (Stockholm, Druck u. Verl. Hasse W. Tullberg, 1904);\\ 
{\it Potr\"atgallerie der Astronomischen Gesellschaft} (Budapest, 
K\"onigliche Ungarische Universit\"atsdruckerei, 1931} 
(Fig.~\ref{fig:portraits}).
His disciple K. Lassovszky wrote a few years after Hark\'anyi's death:

\begin{quote}
While his financial means allowed him to conduct a life free of
worries, he lived a relatively secluded life devoted to scientific
studies. He had a selfless interest in science, more than anybody
else. Extraordinarily well informed in all branches of astronomy, he
also kept a keen interest in the latest developments in theoretical
physics until his last days. He had a very strong critical streak and
he was a particularly stern critic of his own work. He had few
students in the strict sense of the word but those few he let close
enough have never ceased learning from him and deeply regret his
departure.\endnote{K. Lassovszky, ``A magyar csillag\'aszat halottai'', {\it 
Csillag\'aszati Lapok} 1 (2), 1938: pp.~66--67.}
\end{quote}

K\"ovesligethy, who survived Hark\'anyi by two years, gave a necrologue at his
funeral\endnote{R. K\"ovesligethy, ``Besz\'ed b\'ar\'o Hark\'anyi B\'ela l.~tag 
ravatal\'an\'al 1932. janu\'ar 25-\'en'', {\it Akad\'emiai \'Ertes\'{\i}t\H o} 
8, 1932: pp.~79--81 -- 
\url{https://adtplus.arcanum.hu/hu/view/AkademiaiErtesito_1932/?pg=90&layout=s}}, recalling Hark\'anyi's close friendship with Lor\'and E\"otv\"os, their
fellow professor (whose name the university wears today). K\"ovesligethy further
calls Hark\'anyi his ``first disciple, later a colleague and a faithful friend to
the very end. An academic with an honestly modest attitude, avoiding publicity
as much as possible, even in the area of outreach. A man living for scientific
research and lifelong learning''.

\section{Contributions to stellar physics}

Soon after the publication of Wien's blackbody radiation law in 1897 and its
application by Lummer and Pringsheim to temperature determination in laboratory
settings in 1900, Hark\'anyi realized that the same method, essentially consisting
in the fitting of a blackbody radiation curve (according to Wien's formula) to
the measured radiation intensities, could be straightforwardly applied to stars,
provided the necessary spectrophotometric data are available. And, owing to his
broad experience, he was well aware that such data existed:
spectrophotometric data in eight bands had been visually determined at the
Observatory of Potsdam by J. Vogel for five bright stars already in
1880.\endnote{J. Vogel, {\it Monatsberichte der K\"oniglichen Preussische 
Akademie des Wissenschaften zu Berlin,} 1880: pp. 801--811. -- \\
\url{https://archive.org/details/monatsberichtede1880knig/page/800}\\
It is worth noting that the first temperature determination by fitting
a blackbody spectrum was due to K\"ovesligethy (1885, op.cit.~in note
11) who, using his own blackbody formula, determined the Sun's
temperature as 5596 K. For further details see\\
L. G. Bal\'azs, ``Theoretical astrophysics in the 19th century'', {\it
Acta Historica Astronomi{\ae},} 24, 2004: 156--171.} 

Vogel's measurements needed to be corrected for atmospheric extinction. For
this, the zenith distance of the measured objects at the time of the measurement
should have been known. As these data were not at Hark\'anyi's disposal, he was
forced to assume a fiducial value of 45 degrees for the zenith distance (with
the exception of Sirius where its minimal zenith distance of 68{\fdg}9 was
used). This is undoubtedly a major source of error in Hark\'anyi's results.

Following the extinction correction, the measured intensity ratios against the
intensity in the solar spectrum were calculated for each of the 8 wavelengths
and fitted with a blackbody spectrum, one of the fit parameters yielding
($\lambda_{\rm max}-\lambda_{{\rm max,}\odot}$), allowing for a determination of
the temperature relative to the solar temperature from Wien's displacement law.

\begin{figure}
\begin{center}
\includegraphics[width=0.49\textwidth]{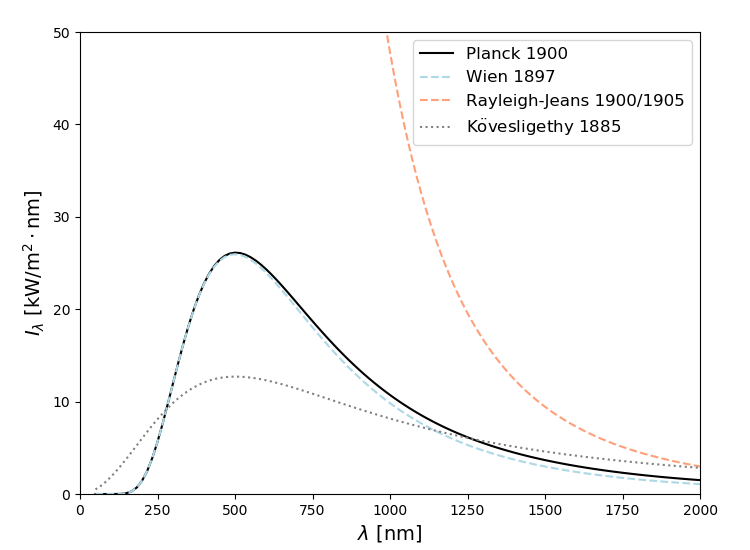}
\includegraphics[width=0.49\textwidth]{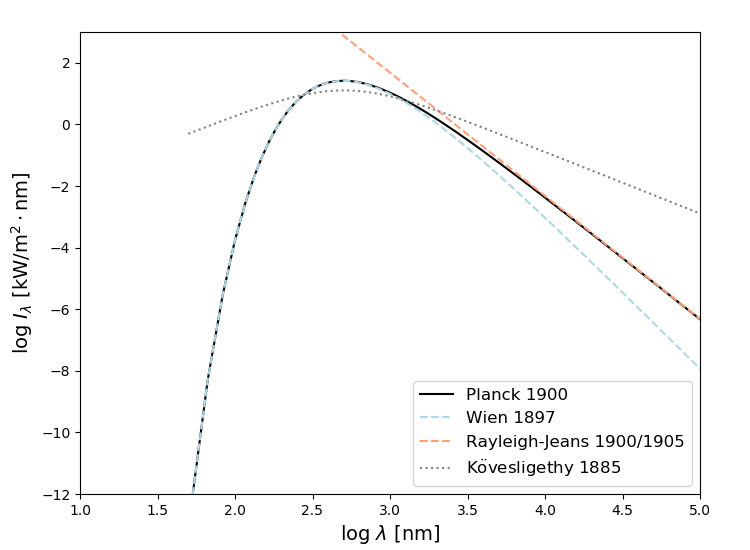}
\end{center}
\caption{Historically suggested energy distributions for blackbody radiation on
a linear scale (left) and on a log--log scale (right).
$T=5770 K$ was assumed for each formula with the currently accepted values of
the physical constants involved.}
\label{fig:bbodies}
\end{figure}

Hark\'anyi's application of the method resulted in the values listed in the second
column of  Table~\ref{table:res}. These temparature values, as published by
Hark\'anyi in 1902, represent the first temperature determinations for individual
stars other than the Sun in the history of astrophysics.

It is apparent from the table that the values obtained were subject to
significant systematic and random errors. Looking for potential sources of these
errors several effects need to be taken in consideration.

\begin{table}
\caption{Surface temperatures of bright stars as determined by Hark\'anyi
(1902), by Wilsing and Scheiner (1910) and by modern observers.}\endnote{U. 
Heiter et al., ``Gaia FGK benchmark stars: Effective temperatures and surface 
gravities''. {\it Astronomy and Astrophysics} 582, 2015: A49;\\ 
T. Ryabchikova, N. Piskunov and D. Shulyak, ``On the Accuracy of Atmospheric
Parameter  Determination in BAFGK Stars'', {\it ASP Conf.~Proc.} 494, 2014:
308;\\ 
R. Monier et al., ``The flat bottomed lines of Vega'', {\it Proceedings of the
Annual meeting of the French Society of Astronomy and Astrophysics,} 2017:
pp.49-55; \\
H. J\"onsson et al. ``Abundances of disk and bulge giants from hi-res optical 
spectra'', {\it Astronomy \& Astrophysics,} 598, 2017: A100; \\
E.~M.~Levesque et
al., ``The Effective Temperature Scale of Galactic Red Supergiants:  Cool, but
Not as Cool as We Thought'', {\it The Astrophysical Journal} 628, 2005:
973--985; \\
H. A. McAlister et al., ``First Results from the CHARA Array. I. An
Interferometric and Spectroscopic Study of the Fast Rotator $\alpha$ Leonis
(Regulus)'', {\it The Astrophysical Journal} 628, 2005: 439-452; \\
R. E. Luck, ``Abundances in the Local Region II: F, G, and K Dwarfs and 
Subgiants'', {\it The Astronomical Journal} 153, 2017: 21.}
\label{table:res}
\begin{center}
\begin{tabular}{lrrrl}
\hline
Object & $T_{\rm H}$ & $T_{\rm WS}$ & $T_m$ & Reference \\
\hline
Sun        &  5450 &	  & 5770  & Heiter et al. 2015, Table 10 \\
Sirius     &  7950 &	  & 9900  & Ryabchikova et al.  2015 \\
$\alpha$ Lyr  &  6400 &	  & 9600  & Monier et al. 2017 \\
$\alpha$ Tau  &  2850 &	  & 3930  & Heiter et al. 2015, Table 10 \\
$\alpha$ Boo  &  2700 & 3500 &  4300 &  J\"onsson et al. 2017 \\
$\alpha$ Ori  &  3150 & 2900 &  3650 &  Levesque et al. 2005, Table 4 \\
$\beta$ Gem   &       & 4400 &  4850 &  J\"onsson et al. 2017 \\
$\alpha$ Leo  &       & 9400 & 12900 &  McAlister et al. 2005 \\
$\alpha$ Aql  &       & 7100 &  7380 &  Luck 2017 \\
$\alpha$ Ari  &       & 3700 &  4460 &  J\"onsson et al. 2017 \\
\hline
\end{tabular}
\end{center}
\end{table}

In his paper submitted in 1901 Hark\'anyi still refrained from the use of
Planck's very recent form of the blackbody spectrum, using Wien's form instead;
as it is also apparent from Fig.~\ref{fig:bbodies}, the error introduced by this
is quite small.  Imprecise knowledge of the constant in Wien's displacement law
(2940 $\mu$m$\cdot$K as opposed to the correct value of 2897) introduces an only
slightly larger error of  1.5\,\%.

A more important source of error is the need for a reliable value of
$\lambda_{{\rm max,}\odot}$. Lacking better experimental data, Hark\'anyi relied
on somewhat outdated measurements, which resulted in a value of 540 nm instead
of the correct 500 nm, introducing a systematic error of $-8\,$\% into the
temperature determinations. 

Correcting for the above effects would increase Hark\'anyi's values by 6.5\,\%,
bringing his values for the Sun and for the artificial light sources (not shown
in our table) into better agreement with reality. For the other stars, however,
the temperature values still remain systematically on the low side even after
this correction. This may point to a potentially important contamination of
Vogler's original measurements by scattered light, which would primarily affect
the fainter sources. 

Hark\'anyi's results made some immediate impact\endnote{K. Bohlin,  {\it Deutsche
Revue} 1902: 12;  S. A. Arrhenius, {\it Worlds in the making; the evolution of
the Universe}  (New York and London, Harper \& brothers, 1908); A. Hnatek,
``Bestimmung einiger effektiver Sterntemperaturen und relativer Sterndurchmesser
auf spektralphotographischem Wege'', {\it Astronomische Nachrichten} 187, 1911:
pp. 369--382.}  but they were
soon surpassed by the work of Wilsing and Scheiner\endnote{J. Wilsing and J.
Scheiner, Temperaturbestimmung von 109 helleren Sternen aus 
spektralphotometrischen Beobachtungen, {\it Astronomische Nachrichten} 183,
1909: pp.~97--108} who collected new spectrophotometric data for 109 bright
stars and determined their temperatures. Their results for the five brightest
stars in their sample are shown in the third column of Table~\ref{table:res} for
comparison. The values are still significantly underestimated, presumably again
due to the influence of scattered light.

Hark\'anyi subsequently realized that a knowledge of the surface temperature
allows the determination of the surface brightness (intensity) of stars in the
visual domain as
\begin{equation}
I_V=\int_{\lambda_1}^{\lambda_2} s(\lambda) B_\lambda(T)\,d\lambda 
\end{equation}
where $\lambda_1$  and $\lambda_1$ are the wavelengths limiting the visual
domain and $B_\lambda(T)$ is the blackbody spectrum and $s(\lambda)$ is the
sensitivity profile of the human eye. $I_V$ had been evaluated analytically by
K\"ovesligethy\endnote{R. de K\"ovesligethy, ``The physical meaning of the
star-magnitude'', {\it The Astrophysical Journal} 11, 1900: pp. 350--356}
in the Wien approximation under the assumption $s=\,$const., and
numerically (graphically) by Hertzsprung\endnote{E. Hertzsprung, ``\"Uber die 
optische St\"arke der Strahlung des schwarzen K\"orpers'', {\it
Zeitschrift f\"ur
wissenschaftliche Photographie, Photophysik und Photochemie} 4, 1906: 43} 
for the Planck function and a wavelength-dependent eye sensitivity. 
{\rev As already realized by Hertzsprung,\endnote{{\rev E. Hertzsprung, ibid.\\ 
Hertzsprung already drew the conclusion that the sizes of some stars
may greatly exceed that of the Sun, i.e. giant stars exist. As a
demonstration, he determined the angular radius of Arcturus as
0{\farcs}25. Hark\'anyi's value for the same star was 0{\farcs}19,
while the modern value is close to 0{\farcs}1. For further discussion of 
the history of stellar diameter determinations see\\ 
D.~H.~DeVorkin, ``Michelson and the Problem of Stellar Diameters'', 
{\it Journal for the History of Astronomy} 6, 1975: pp.~1--18}}}
Pogson's formula then allows to convert intensity to surface
magnitude, and a division of the apparent visual magnitude with the
surface magnitude yields the apparent angular size of the stellar
disk. The physical diameter of the star may be determined from this if
the distance is known. This programme was carried out by Hark\'anyi in
his studies of 1910, resulting in apparent angular radii derived for
156 stars. These results represented a significant improvement over
the pioneering work of Pickering thirty years earlier, where a uniform
surface brightness was assumed for single stars and a better
approximation was only possible for binaries.\endnote{E. C. Pickering, 
``Dimensions of the fixed stars  with especial reference to binaries
and variables of the Algol type'',  {\it Proceedings of the American
Academy of Arts and Sciences} 16, 1880:  pp. 1--37}.

\section{Hark\'anyi's place in the history of astrophysics}

Hark\'anyi was, beyond doubt, the first to determine surface temperature values
for individual stars other than the Sun. {\rev Following Hertzsprung's
pioneering studies,} he was also the first to determine the
apparent radii of {\rev a significant number of} single stars on physically correct grounds with the radiation
energetics method. His relatively obscure status in the history of science may
be ascribed to a combination of his modest and reclusive character and his low
productivity. Even with the standards of an age when single-authored papers were
the norm and ``publish or perish'' did not yet prevail, the quantity of
Hark\'anyi's scientific output was low. The Great War and its aftermath resulted
in a further decline in his research activity and perhaps his motivation, but he
has always remained one of those scientists who do not publish often but when
they do, it deserves attention.

The trail opened by Hark\'anyi towards the determination of stellar parameters
has developed into a broad multi-lane highway, the latest major extension of
which has been Gaia Data Release 2. The ``effective'' temperature values in 
Gaia DR2 are actually based on multicolor photometry (used in combination with a
machine learning algorithm). The r.m.s.\ error resulting from this procedure is
estimated to be 324 K, which is quite significant: the $2\sigma$ error interval
covers a range of nearly 1300 K. Keeping in mind, however, that the sample
is mostly comprised of stars of the 17th magnitude, the result is quite
satisfactory. 

As with Hark\'anyi's pioneering work, extinction is a main source of uncertainty
in the Gaia temperature estimates, but this time it is interstellar rather than
atmospheric extinction. Some studies based on Gaia temperatures found that
in subsets of the Gaia sample selected according to physical criteria $T_{\rm
eff}$ values can be subject to a heavy systematic bias.\endnote{A. B\'odi and L.
L. Kiss, ``Physical Properties of Galactic RV Tauri Stars from Gaia DR2 Data'',
{\it The Astrophysical Journal} 872, 2019: 60.}

The near coincidence of the 150th anniversary of Hark\'anyi's birth with the the
unprecedented increase in the number of stars with parameter estimates from Gaia
DR2, together with the distant parallelisms in the limitations affecting both
works provided the motivation for this short essay.

\begin{acks}
The author has greatly profited from enlightening discussions with Dr. L. G.
Bal\'azs.
\end{acks}

\begin{biogs}
Prof.~Krist\'of Petrovay is the Head of Department of Astronomy at E\"otv\"os
Lor\'and University (ELTE), Budapest, Hungary, where he obtained
tenure following a graduate and postdoctoral research career spent at ELTE, at
the University of Oxford and at  the Instituto de Astrof\'{\i}sica de Canarias.
His main research field is solar magnetohydrodynamics and space climate. His
interest in the history of astronomy focuses on aspects related to the 300 year
long history of astronomy at ELTE.
\end{biogs}

\theendnotes

\end{document}